\documentclass{PoS}

\title{Flavour-violation in two-Higgs-doublet models}
\usepackage{amsmath}
\usepackage{amssymb}
\ShortTitle{2HDM III}

\author{\speaker{Andreas Crivellin}, Christoph Greub and Ahmet Kokulu, 
\thanks{This work is supported by the Swiss National Science Foundation (SNF). Support from the Funds for the promotion of young scientists of the 
Faculty of Sciences is gratefully acknowledged.}\\
        Albert Einstein Center for Fundamental Physics, \\Institute for Theoretical Physics, University of Bern.\\
        E-mail: \email{crivellin@itp.unibe.ch}, \email{greub@itp.unibe.ch}, \email{akokulu@itp.unibe.ch}}

\abstract{In these proceedings we review the flavour phenomenology of 2HDMs with generic Yukawa structures \cite{Crivellin:2013wna}. We first consider the quark sector and find that despite the stringent constraints from FCNC processes large effects in tauonic $B$ decays are still possible. We then consider lepton flavour observables, show correlations between $\mu\to e\gamma$ and $\mu^-\to e^-e^+e^-$ in the 2HDM of type III and give upper bounds on the lepton flavour violating $B$ decay $B_{d}\to \mu e$.}

\FullConference{The European Physical Society Conference on High Energy Physics -EPS-HEP2013\\
		18-24 July 2013\\
		Stockholm, Sweden}

\begin{document}

\section{Introduction}

\begin{figure}[htbp]
\begin{center}
\includegraphics[width=0.49\textwidth]{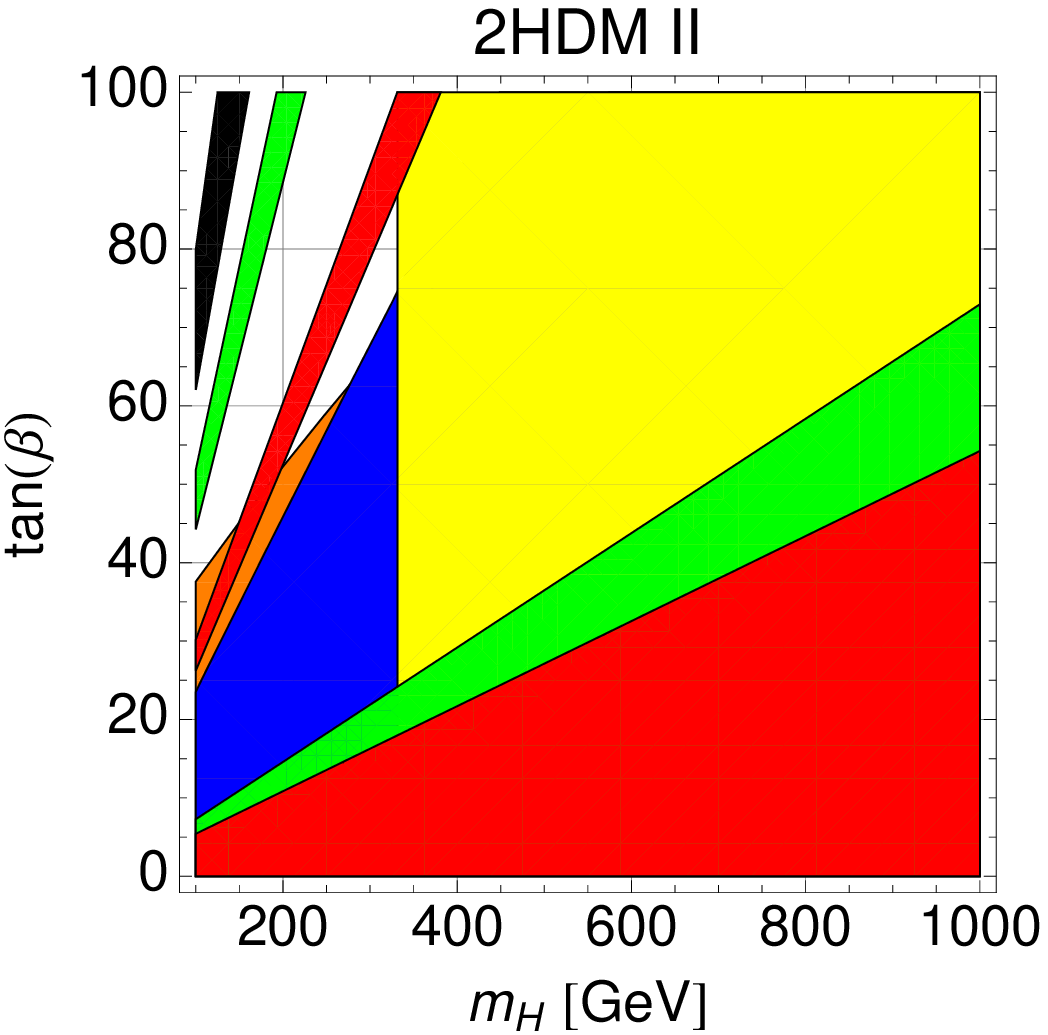}
\includegraphics[width=0.5\textwidth]{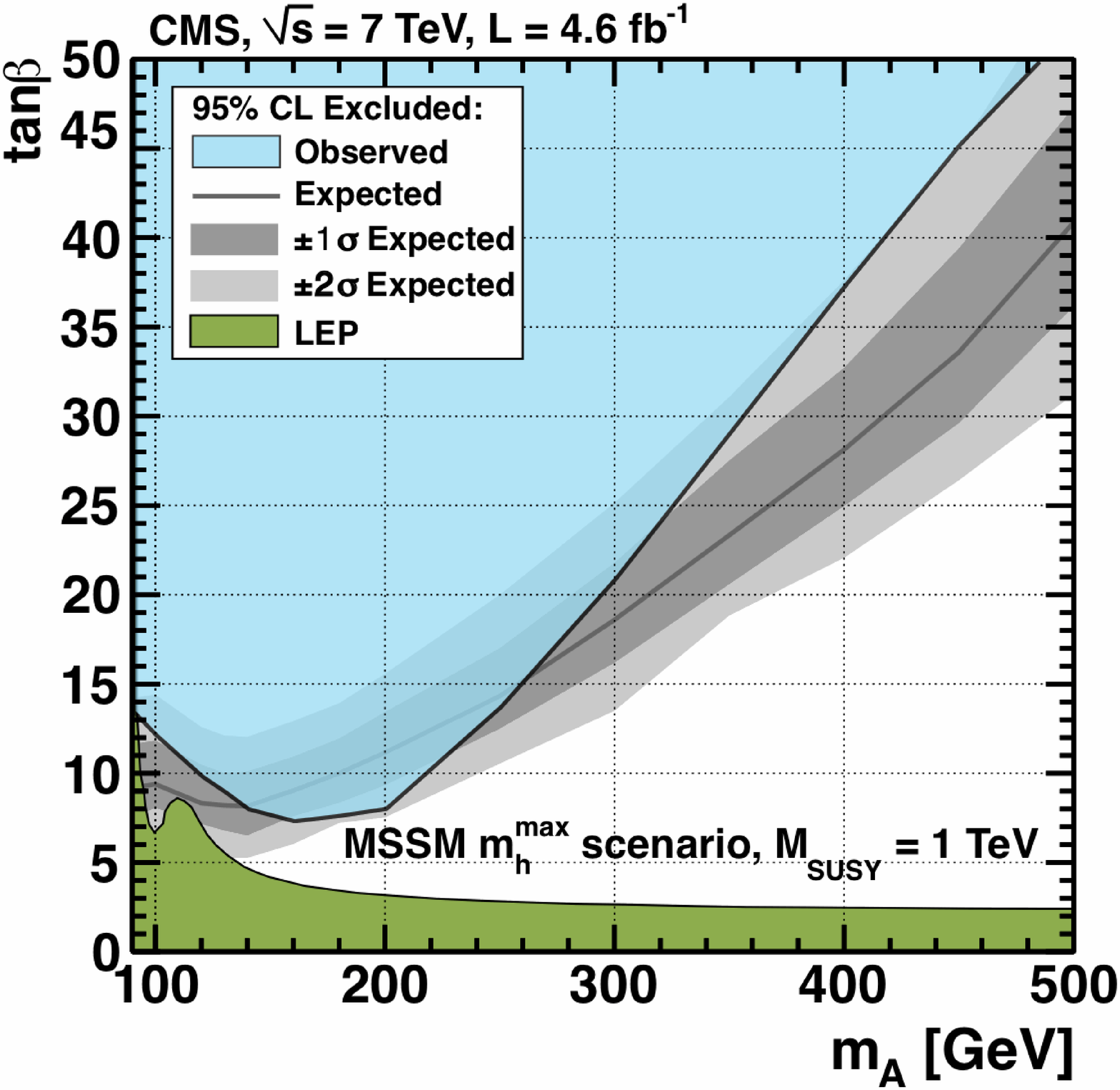}
\end{center}
\caption{Left: Updated constraints on the 2HDM of type II parameter space. The regions compatible with experiment are shown (the regions are superimposed on each other): $b\to s\gamma$ (yellow) \cite{Hermann:2012fc}, $B\to D\tau\nu$ (green), $B\to \tau \nu$ (red), $B_{s}\to \mu^{+} \mu^{-}$ (orange), $K\to \mu \nu/\pi\to \mu \nu$ (blue) and $B\to D^*\tau \nu$ (black). Note that no region in parameter space is compatible with all processes. Explaining $B\to D^*\tau \nu$ would require very small Higgs masses and large values of $\tan\beta$ which is not compatible with the other observables. To obtain this plot, we added the theoretical uncertainty linear on the top of the $2 \, \sigma$ experimental error.\newline
Right: Plot from the CMS collaboration taken from Ref.~\cite{CMS}: Exclusion limits in the $m_{A^0}$--$\tan\beta$ plane from $A^{0}\to \tau^{+}\tau^{-}$. The analysis was done in the MSSM, but since we consider a 2HDM with MSSM-like Higgs potential and the MSSM corrections to the $A^0\tau\tau$ vertex are small, we can apply this bound to our model. However, a large value of $\epsilon^\ell_{33}$ in the 2HDM of type~III could affect the conclusions. Note that in the limit $v\ll m_H$ all heavy Higgs masses ($m_{H^0}$, $m_{A^0}$ and $m_{H^\pm}$) are approximately equal.}
\label{fig:2HDMII}
\end{figure}

The SM contains only one scalar isospin doublet, the Higgs doublet. After electroweak symmetry breaking, this gives masses to up quarks, down quarks and charged leptons. The charged component of this doublet becomes the longitudinal component of the $W$ boson and the neutral CP-odd component becomes the longitudinal component of the $Z$ boson. Thus we have only one physical neutral Higgs particle. In a 2HDM \cite{Lee:1973iz} we introduce a second Higgs doublet and obtain four additional physical Higgs particles: the neutral CP-even Higgs $H$, a neutral CP-odd Higgs $A$ and the two charged Higgses $H^{\pm}$ (in the case of a CP conserving Higgs potential). The most general Lagrangian for the Yukawa interactions (which corresponds to the 2HDM of type III) in the physical basis with diagonal quark mass matrices is given by
\begin{eqnarray}
\mathcal{L}_Y = &-& \bar d_{f\,L} \left[\left(\dfrac{{m_{d_i }
  }}{{v_d }}\delta_{fi} - \epsilon_{fi}^{ d}\tan\beta
  \right)H_d^{0\star}\,+\,\epsilon_{fi}^{ d}\,H_u^0 \right]d_{i\,R}
\nonumber \\
&-& \bar u_{f\,L} \left[\left(\dfrac{{m_{u_i } }}{{v_u }}\delta_{fi} -
  \epsilon_{fi}^{ u}\cot\beta \right)H_u^{0\star}\,+\,\epsilon_{fi}^{ u}\,H_d^{0} \right] u_{i\,R} \nonumber\\
&+& \bar u_{f\,L} V_{fj} \left[ {\dfrac{{m_{d_i } }}{{v_d
    }}\delta_{ji}-\left( {\cot \beta + \tan \beta }
    \right) \epsilon_{ji}^{ d}  } \right]H^{2\star}_d\ d_{i\,R} \nonumber \\
&+& \bar d_{f\,L} V_{jf}^{\star} \left[ { \dfrac{{m_{u_i
    } }}{{v_u }}\delta_{ji}-\left( {\tan \beta +
      \cot\beta } \right)\epsilon_{ji}^{ u}  } \right] H^{1\star}_u u_{i\,R}\,+\,{h.c.}  \,\,\, .
\label{L-Y-FCNC}
\end{eqnarray}
where $\epsilon^q_{ij}$ parametrizes the non-holomorphic corrections which couple up (down) quarks to the down (up) type Higgs doublet\footnote{The expression for the charged leptons where $\epsilon^\ell_{ij}$ parametrizes the non-holomorphic corrections can be obtained from the one for down-quarks by setting substituting $\ell$ for $d$ and setting $V_{ij}=\delta_{ij}$ (for massless neutrinos).}. In the MSSM at tree-level $\epsilon^q_{ij}=0$, which also corresponds to the 2HDM of type II, and flavour changing neutral Higgs couplings are absent. A combination of flavour constraints on the 2HDM of type II is given in the left plot of Fig.~\ref{fig:2HDMII}.

However, at the loop-level, the non-holomorphic couplings $\epsilon^q_{ij}$ are generated~\cite{Hamzaoui:1998nu}\footnote{See the second article of Ref.~\cite{Crivellin:2011jt} for a complete treatment of all chirally enhanced effects.} and in the following we will assume that $\epsilon^q_{ij}$ are free parameters but are small corrections compared to the Yukawa coupling.

\section{Quark flavour physics}

\subsection{Constraints from FCNC processes}

\begin{figure*}[t]
\centering
\includegraphics[width=0.3\textwidth]{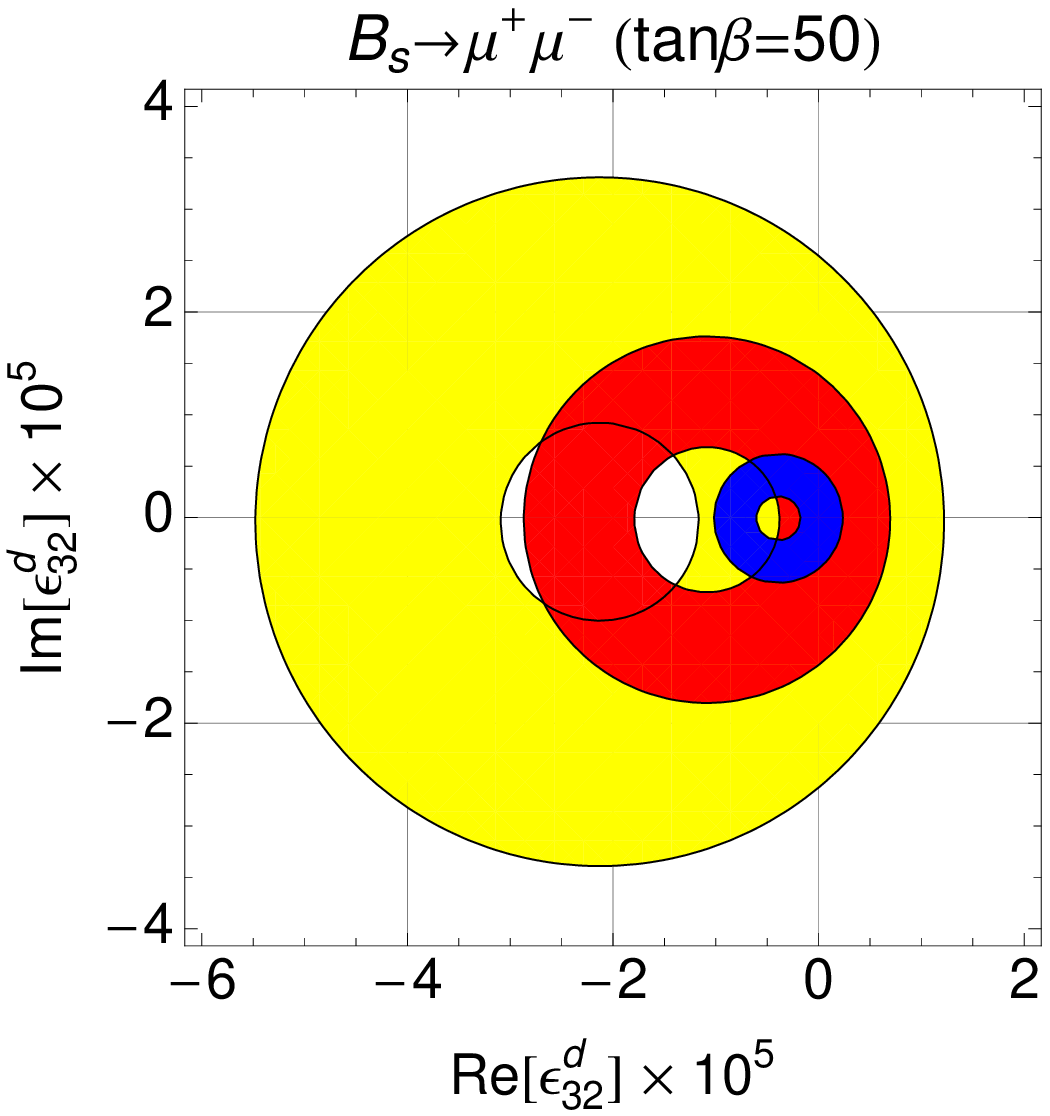}~~~~~
\includegraphics[width=0.3\textwidth]{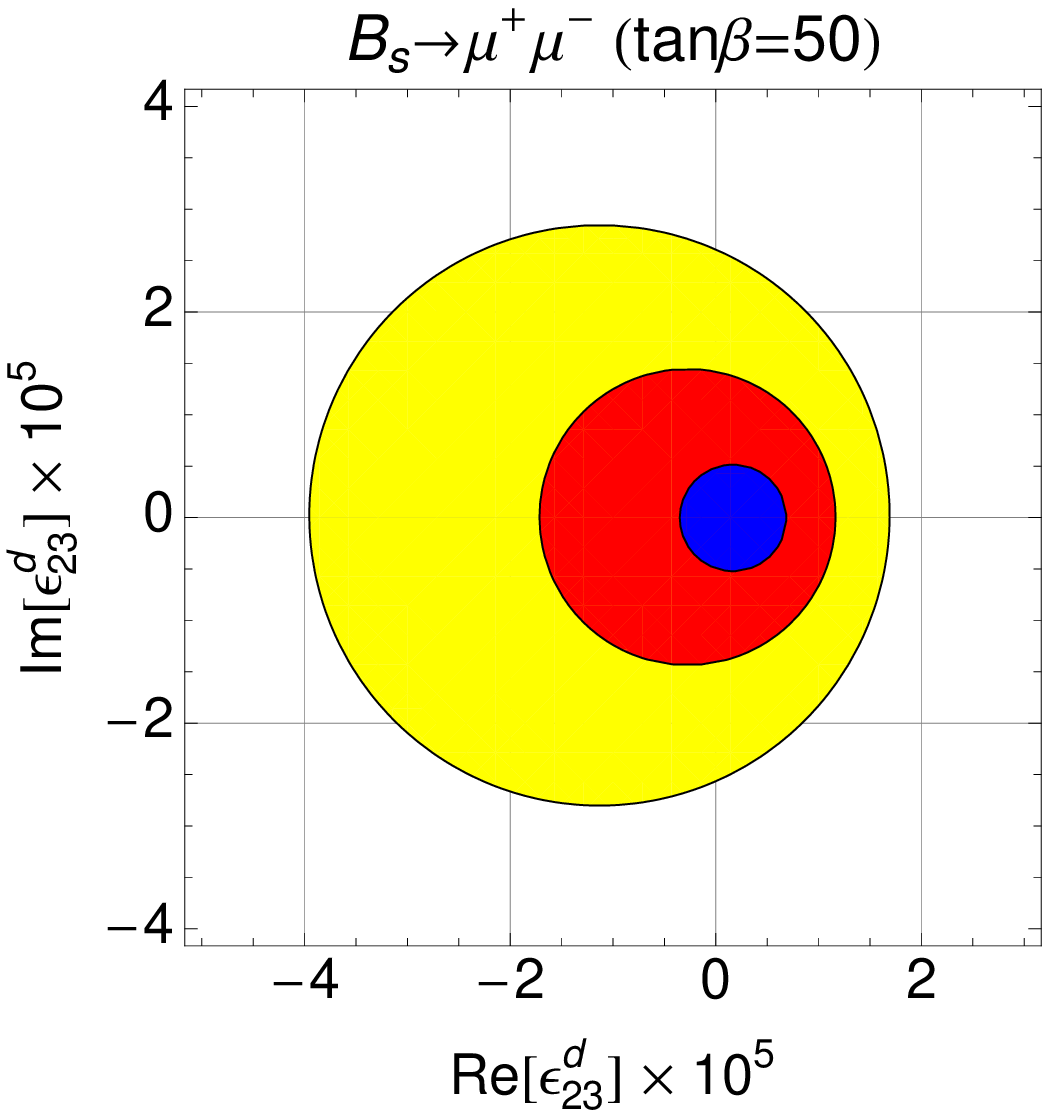}
\includegraphics[width=0.334\textwidth]{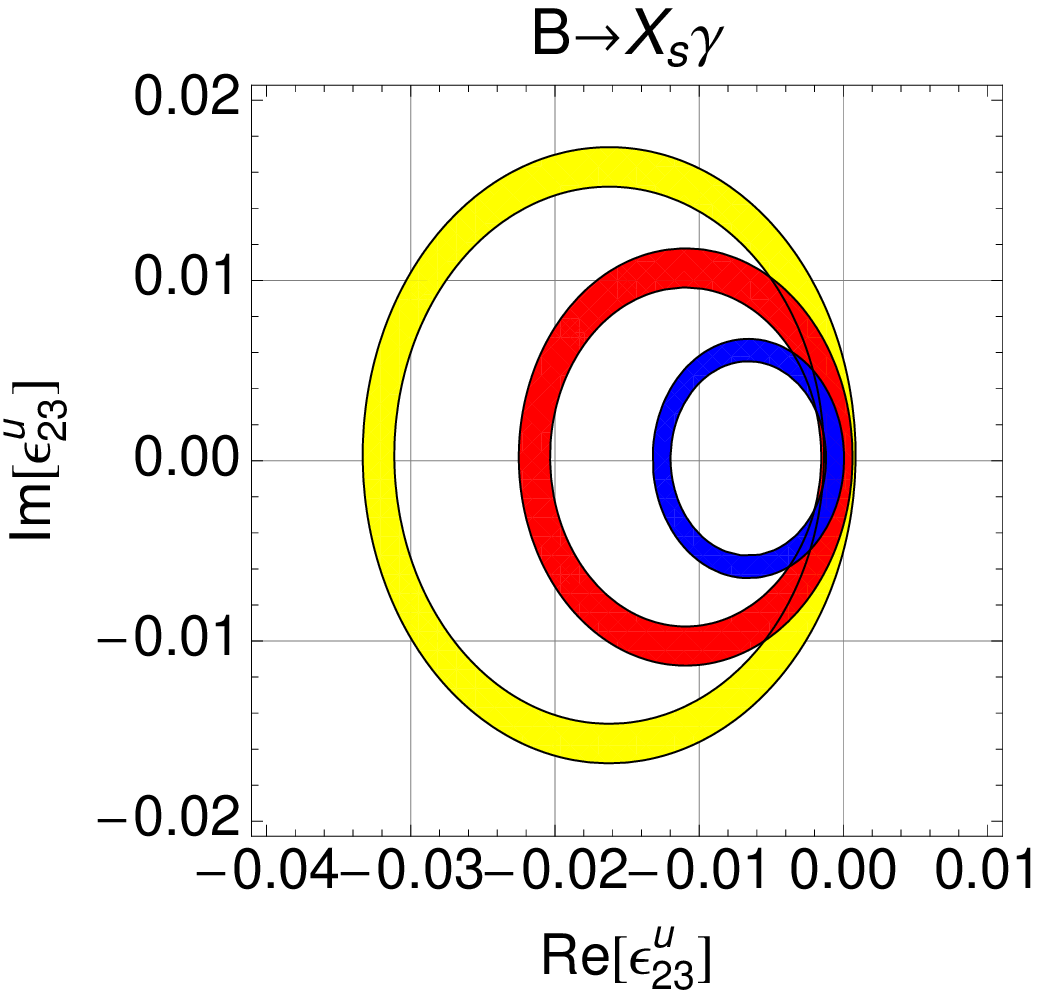}
\caption{Left (middle): Allowed regions in the complex $\epsilon^{d}_{23(32)}$--plane from $B_s\to\mu^+\mu^-$ for $\tan\beta=50$ and $m_{H}=700\mathrm{~GeV}$ (yellow), $m_{H}=500\mathrm{~GeV}$ (red) and $m_{H}=300\mathrm{~GeV}$ (blue). Note that the allowed regions for $\epsilon^{d}_{32}$--plane are not full circles because in this case a suppression of ${\cal B}\left[B_{s}\to\mu^+\mu^-\right]$ below the experimental lower bound is possible.\newline
Right: Allowed regions for $\epsilon^{u}_{23}$ from $ B \to X_{s} \gamma$, obtained by adding the $2\,\sigma$ experimental error and theoretical uncertainty linear for $\tan\beta=50$ and $m_{H}=700 \, \mathrm{ GeV}$ (yellow), $m_{H}=500\, \mathrm{ GeV}$ (red) and  $m_{H}=300 \,\mathrm{ GeV}$ (blue). }
\label{fig:Bstomumu}
\end{figure*}

Direct constraints on the off-diagonal elements $\epsilon^q_{fi}$ can be obtained from neutral Higgs contributions to the leptonic neutral meson decays ($B_{s,d}\to\mu^+\mu^-$, $K_L\to\mu^+\mu^-$ and ${\bar D}^0\to\mu^+\mu^-$) which arise already at the tree level\footnote{In principle, the constraints from these processes could be weakened, or even avoided, if $\epsilon^\ell_{22}\approx m_{\ell_{2}}/v_u$. Anyway, in here we will assume that the Peccei Quinn breaking for the leptons is small and neglect the effect of $\epsilon^\ell_{22}$ in our numerical analysis for setting limits on $\epsilon^q_{ij}$.
}. $K_L\to\mu^+\mu^-$ constrains $\left|\epsilon^d_{12,21}\right|$, $D^0\to\mu^+\mu^-$ imposes bounds on $\left|\epsilon^u_{12,21}\right|$ and $B_s\to\mu^+\mu^-$ ($B_d\to\mu^+\mu^-$) limits the possible size of $\left|\epsilon^d_{23,32}\right|$ $\left(\left|\epsilon^d_{13,31}\right|\right)$. We find the following (approximate) bounds on the absolute value of $\epsilon^q_{ij}$:
\begin{equation}
\begin{array}{l}
\left|\epsilon^d_{12,21}\right|\leq 1.6\times 10^{-6}\,,\qquad
\left|\epsilon^u_{12,21}\right|\leq 3 \times 10^{-2}\,,\\
\left|\epsilon^d_{23,32}\right|\leq 3 \times 10^{-5}\,,\qquad
\left|\epsilon^d_{13,31}\right|\leq 1 \times 10^{-5}\,,\\
\end{array}
\end{equation}
for $\tan\beta=50$ and $m_H=500$~GeV. As an example we show the full dependence of the constraints in the complex $\epsilon^{d}_{23,32}$-plane from $B_s  \to\mu^+\mu^-$ in left and middle plot of Fig.~\ref{fig:Bstomumu}. Note that both an enhancement or a suppression of ${\cal B}\left[ B_{d,s}\to\mu^+\mu^-\right]$ compared to the SM prediction is possible. If at the same time both elements $\epsilon^{d}_{23}$ and $\epsilon^{d}_{32}$ are non-zero, constraints from $B_s$ mixing arise which are even more stringent.
\medskip

So far we were able to constrain all flavour off-diagonal elements $\epsilon^d_{ij}$ and $\epsilon^u_{12,21}$ but no tree-level constraints on $\epsilon^u_{13,31}$ and $\epsilon^u_{23,32}$ can be obtained due to insufficient experimental data for top FCNCs. Nonetheless, it turns out that also the elements $\epsilon^u_{13,23}$ can be constrained from charged Higgs contributions to the radiative $B$ decay $b\to d \gamma$ and $ b\to s \gamma$. As an example we show the constraints on $\epsilon^u_{23}$ in the right plot of Fig.~\ref{fig:Bstomumu}. The constraints on $\epsilon^u_{13}$ from $ b\to d \gamma$ are even more stringent \cite{Crivellin:2011ba}.

However, the elements $\epsilon^u_{32,31}$ cannot be seriously constrained from FCNC processes because of the light charm or up quark propagating in the loop (which also requires the contribution to be proportional to this small mass). This has important consequences for charged current processes (to be studied in the next subsection) where these elements enter.
\medskip

\subsection{Tauonic $B$ decays}

Tauonic $B$-meson decays are an excellent probe of new physics: they test lepton flavor universality satisfied in the Standard Model (SM) and are sensitive to new particles which couple proportionally to the mass of the involved particles (e.g. Higgs bosons) due to the heavy $\tau$ lepton involved. Recently, the BABAR collaboration performed an analysis of the semileptonic $B$ decays $B\to D\tau\nu$ and $B\to D^*\tau\nu$ using the full available data set \cite{BaBar:2012xj}. They find for the ratios
\begin{equation}
{\cal R}(D^{(*)})\,=\,{\cal B}(B\to D^{(*)} \tau \nu)/{\cal B}(B\to D^{(*)} \ell \nu)\,,
\end{equation}
the following results:
\begin{eqnarray}
{\cal R}(D)\,=\,0.440\pm0.058\pm0.042  \,,\qquad
{\cal R}(D^*)\,=\,0.332\pm0.024\pm0.018\,.
\end{eqnarray}
Here the first error is statistical and the second one is systematic. Comparing these measurements to the SM predictions
\begin{eqnarray}
{\cal R}_{\rm SM}(D)\,=\,0.297\pm0.017\, \qquad
{\cal R}_{\rm SM}(D^*) \,=\,0.252\pm0.003 \,,
\end{eqnarray}
we see that there is a discrepancy of 2.2\,$\sigma$ for $\cal{R}(D)$ and 2.7\,$\sigma$ for $\cal{R}(D^*)$ and combining them gives a $3.4\, \sigma$ deviation from the SM~\cite{BaBar:2012xj}. This evidence for new physics in $B$-meson decays to taus is further supported by ${\cal B}[B\to \tau\nu]=(1.15\pm0.23)\times 10^{-4}$, which disagrees with the SM prediction by $1.6\, \sigma$ using $V_{ub}$ from a global fit of the CKM matrix \cite{Charles:2004jd}. A natural possibility to explain these enhancements compared to the SM prediction is a charged scalar particle which couples proportionally to the masses of the fermions involved in the interaction: a charged Higgs boson. A charged Higgs  affects $B\to \tau\nu$~\cite{Hou:1992sy}, $B\to D\tau\nu$ and $B\to D^*\tau\nu$~\cite{Tanaka:1994ay}. 

In a 2HDM of type II (with MSSM like Higgs potential) the only free additional parameters are $\tan\beta=v_u/v_d$ (the ratio of the two vacuum expectation values) and the charged Higgs mass $m_{H^\pm}$ (the heavy CP even Higgs mass $m_{H^0}$ and the CP odd Higgs mass $m_{A^0}$ can be expressed in terms of the charged Higgs mass and differ only by electroweak corrections). In this setup the charged Higgs contribution to $B\to \tau\nu$ interferes necessarily destructively with the SM \cite{Hou:1992sy}. Thus, an enhancement of ${\cal B}\left[B\to \tau\nu\right]$ is only possible if the absolute value of the charged Higgs contribution is bigger than two times the SM one\footnote{Another possibility to explain $B\to \tau\nu$ is the introduction of a right-handed $W$-coupling \cite{Crivellin:2009sd}.}. Furthermore, a 2HDM of type II cannot explain ${\cal{R}}(D)$ and ${\cal{R}}(D^*)$ simultaneously \cite{BaBar:2012xj} (as can be also seen in Fig.~1).

\begin{figure*}[t]
\centering
\includegraphics[width=0.3\textwidth]{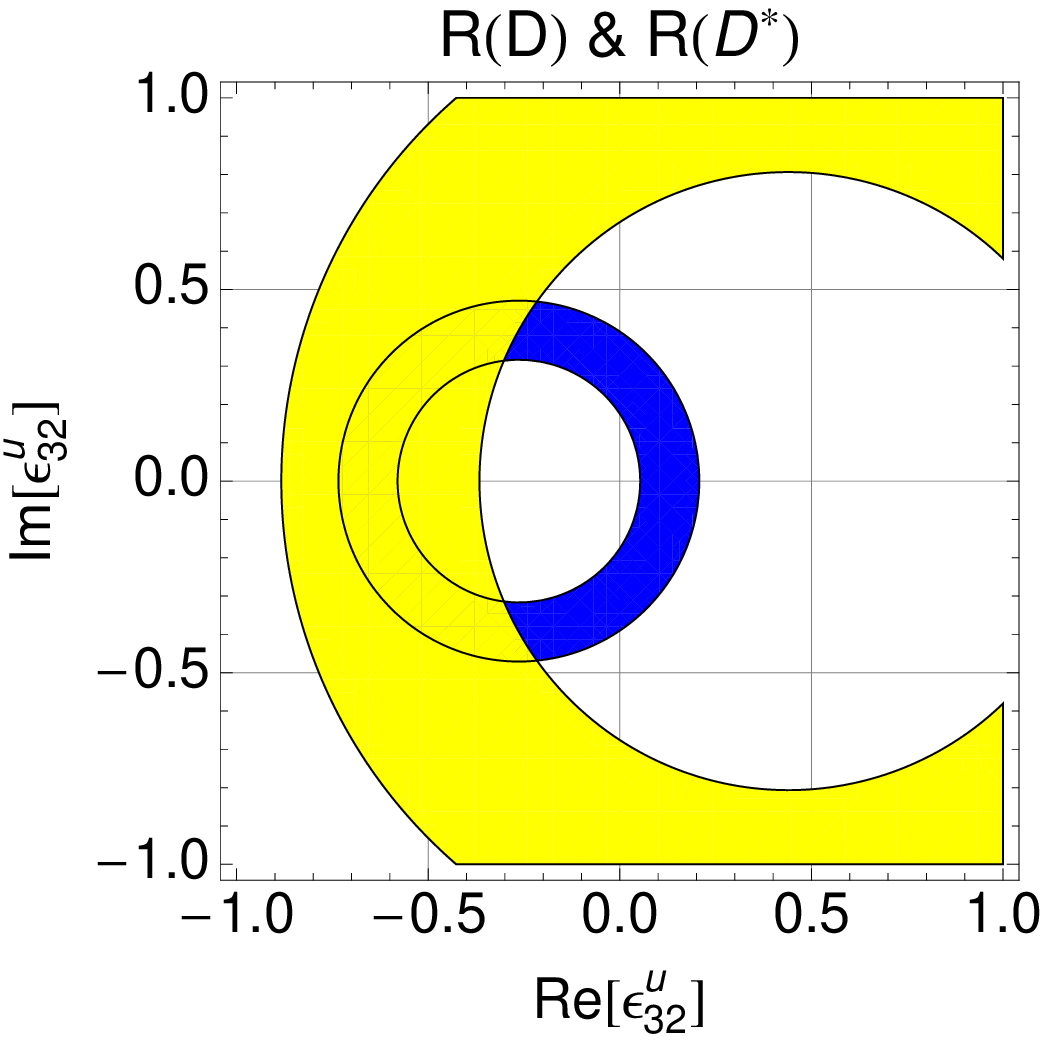}
\includegraphics[width=0.31\textwidth]{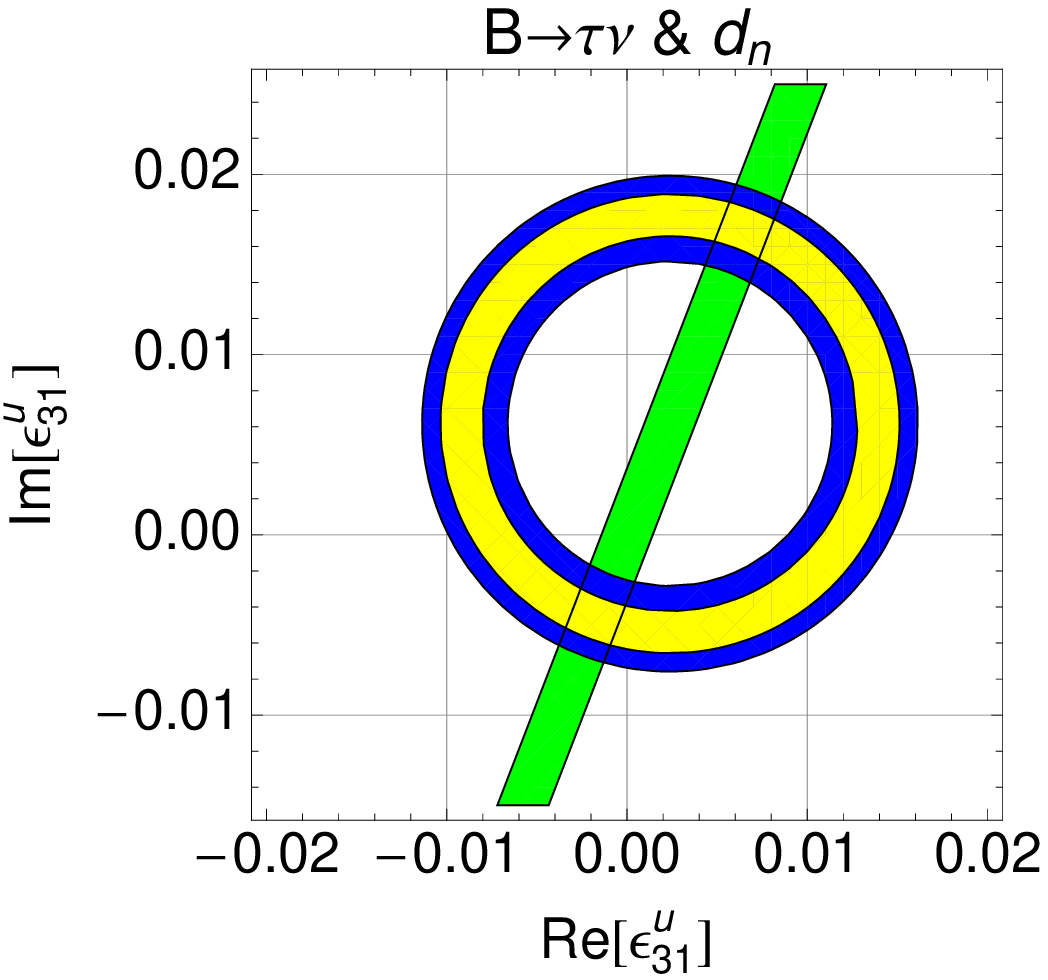}
\includegraphics[width=0.31\textwidth]{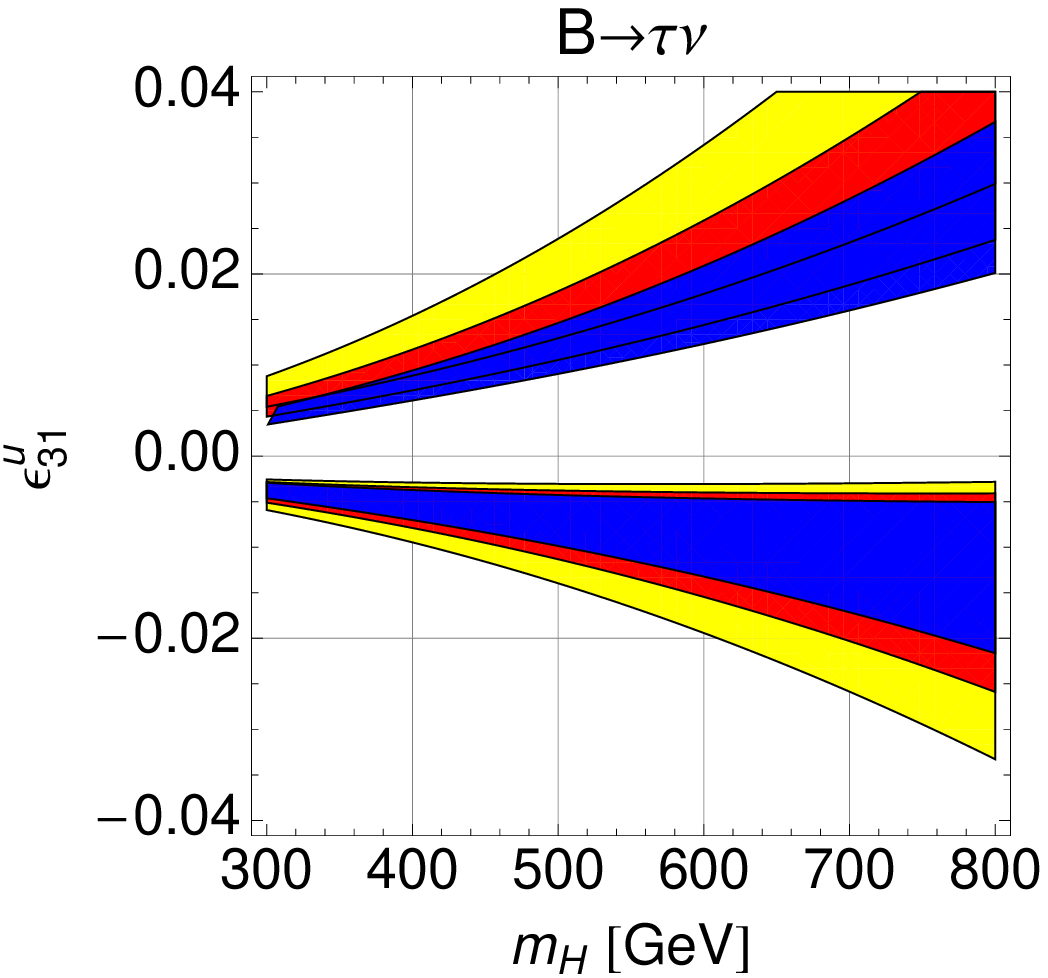}
\caption{Left: Allowed regions in the complex $\epsilon^u_{32}$--plane from ${\cal{R}}(D)$ (blue) and ${\cal{R}}(D^*)$ (yellow) for $\tan\beta=50$ and $m_H=500$~GeV. Middle:  Allowed regions in the complex $\epsilon^u_{31}$--plane combining the constraints from $B \to \tau\nu$ (1 $\sigma$ (yellow) and 2 $\sigma$ (blue)) and neutron EDM (green) for $\tan\beta=50$ and $m_H=500$~GeV. Right: Allowed regions in the $m_{H}$--$\epsilon^u_{31}$ plane from $B \to \tau\nu$ for real values of $\epsilon^u_{31}$ and $\tan\beta=50$~(yellow), $\tan\beta=40$~(red) and $\tan\beta=30$~(blue). $\epsilon^u_{32}$ and $\epsilon^u_{31}$ are given at the matching scale $m_H$. \label{2HDMIII}}
\end{figure*}

We know from the last section that in the 2HDM of type III only $\epsilon^d_{33}$ and $\epsilon^u_{31,32}$ can have sizable effects. 
Indeed, it turns out that by using $\epsilon^u_{32}$ we can explain ${\cal{R}}(D^*)$ and ${\cal{R}}(D)$ simultaneously, while this is not possible with $\epsilon^d_{33}$ \cite{Crivellin:2012ye}. In Fig.~\ref{2HDMIII} we see the allowed region in the complex $\epsilon^u_{32}$-plane, which gives the correct values for ${\cal{R}}(D)$ and ${\cal{R}}(D^*)$ within the $1\, \sigma$ uncertainties for $\tan\beta=50$ and $M_H=500$~GeV. Similarly, $B\to \tau\nu$ can be explained by using $\epsilon^u_{31}$.

\section{Lepton-flavor-violating decays: $\mu\to e\gamma$, $\mu \to e^-e^+e^-$ and $B_d\to\mu e$}
\label{taumumumu}

In this section, we investigate the constraints on the off-diagonal elements $\epsilon^\ell_{ij}$ and study correlations among lepton flavour-violating observables.

\begin{figure}[t]
\centering
\includegraphics[width=0.4\textwidth]{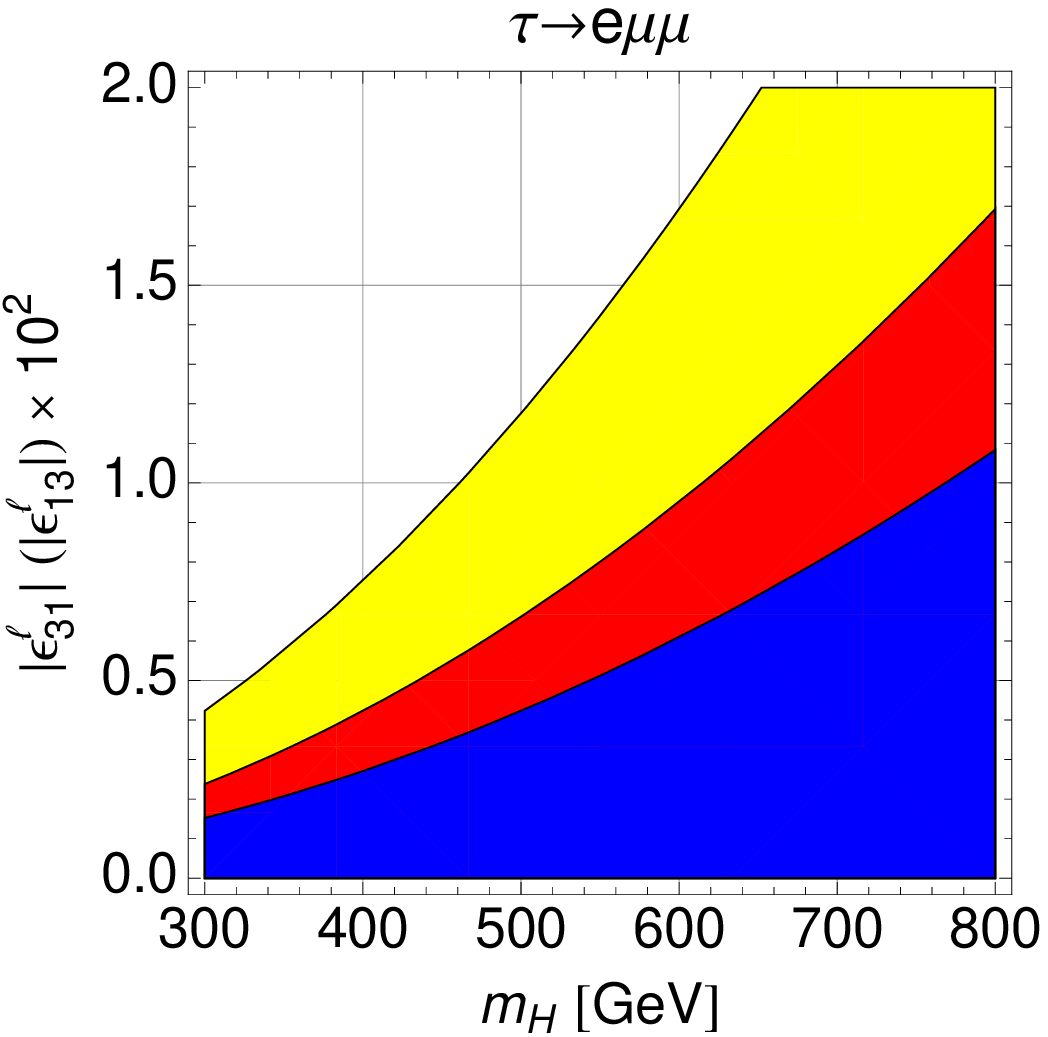}
\includegraphics[width=0.4\textwidth]{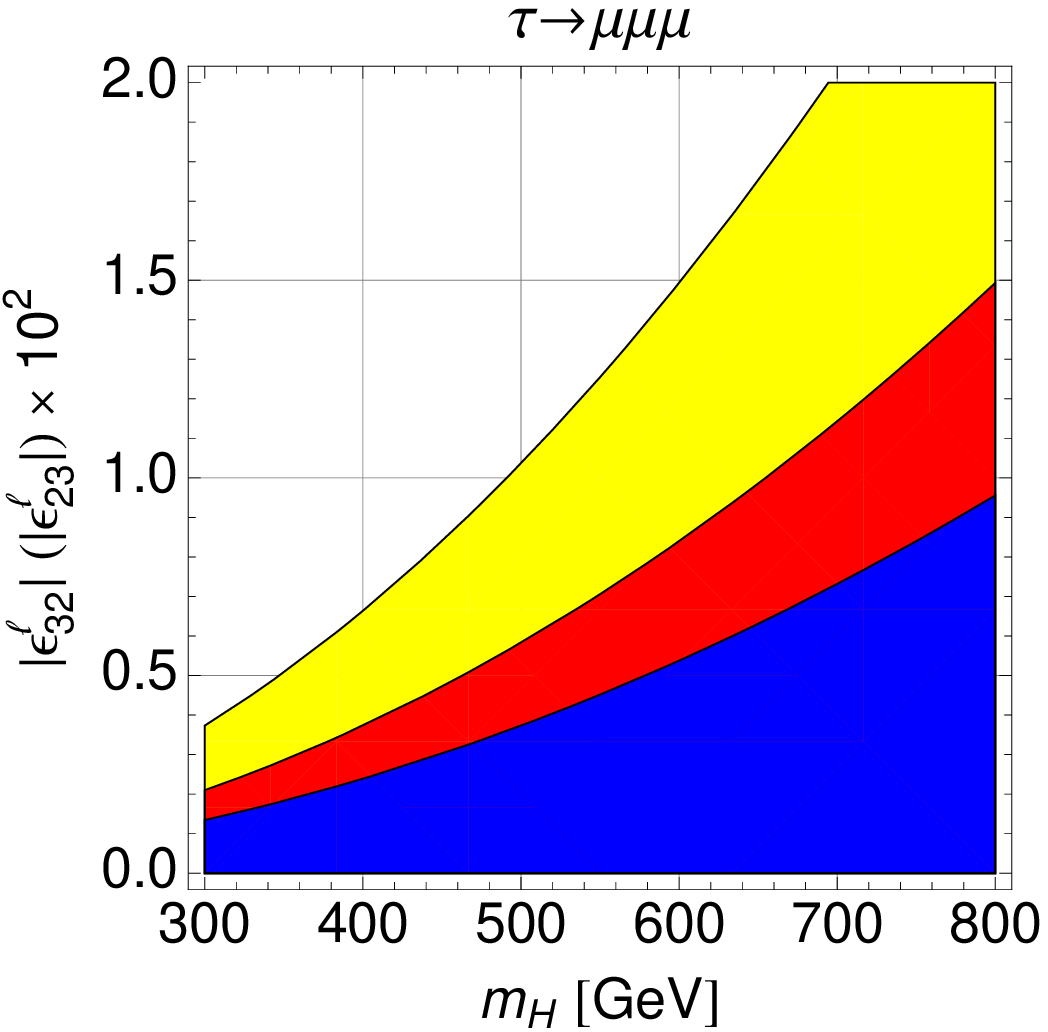}
\caption{ Allowed regions for the absolute value of $\epsilon^{\ell}_{13,31}$ and $\epsilon^{\ell}_{23,32}$ for $\tan\beta=30$ (yellow), $\tan\beta=40$ (red) and $\tan\beta=50$ (blue) from $\tau^-\to e^-\mu^+\mu^-$, $\tau^-\to \mu^-\mu^+\mu^-$ and $\mu^-\to e^-e^+e^-$, respectively. In each plot only one of the elements $\epsilon^{\ell}_{if}$ or $\epsilon^{\ell}_{fi}$ is assumed to be different from zero.}
\label{Fig:tau3mu}
\end{figure}

\begin{figure}[t]
\centering
\includegraphics[width=0.4\textwidth]{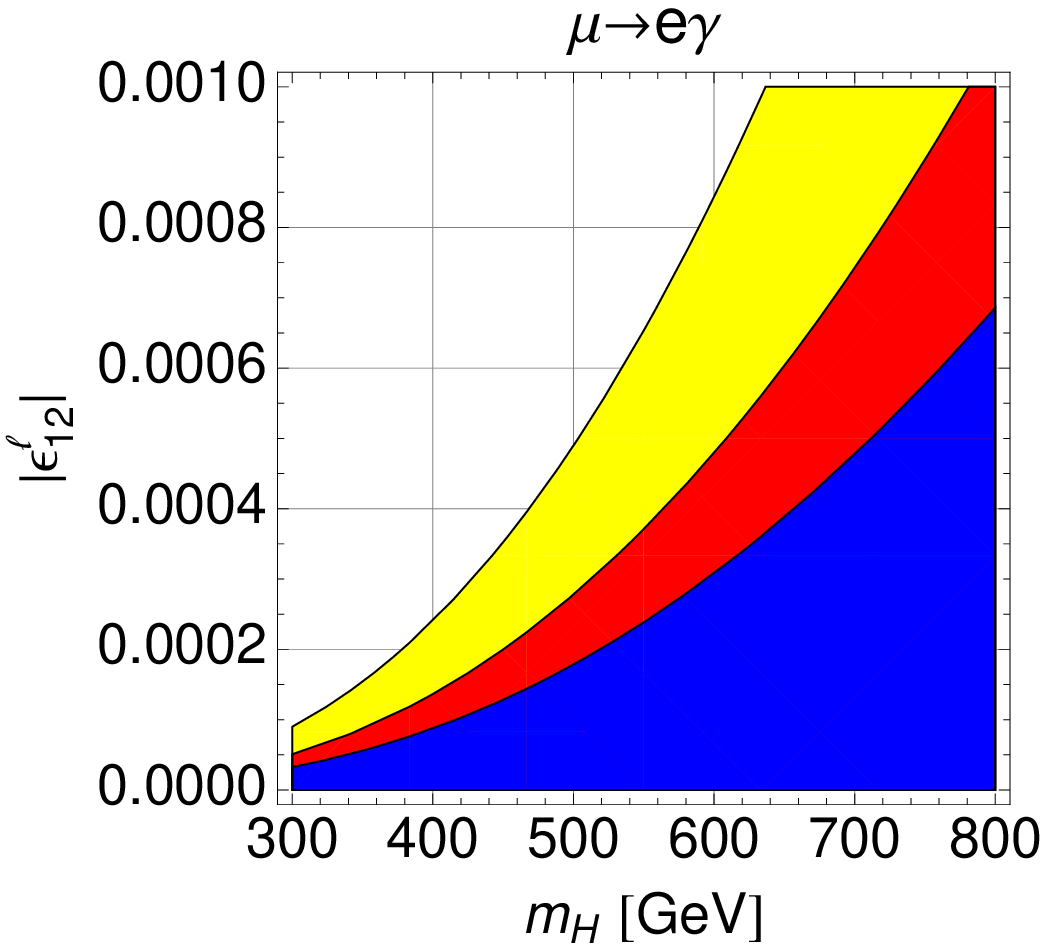}
\includegraphics[width=0.4\textwidth]{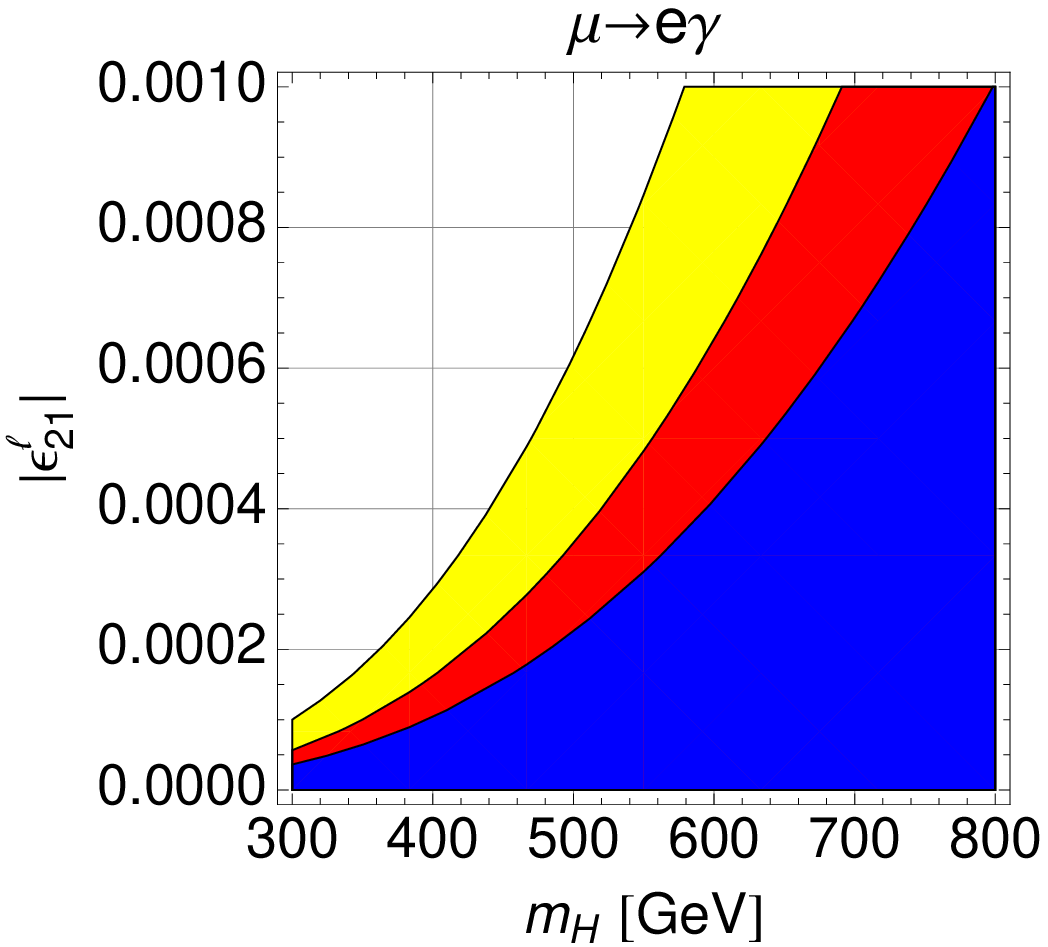}
\caption{ Allowed region for $\epsilon^{\ell}_{12}$ (left plot) and $\epsilon^{\ell}_{21}$ (right plot) from $\mu\to e\gamma$ for $\tan\beta=30$ (yellow), $\tan\beta=40$ (red) and $\tan\beta=50$ (blue).}
\label{Fig:muegamma}
\end{figure}

The bounds on $\epsilon^\ell_{13,31}$ and $\epsilon^\ell_{23,32}$ from the radiative lepton decays $\tau\to e \gamma$ and $\tau\to \mu \gamma$ turn out to be significantly weaker than the ones from $\tau^- \to \mu^-\mu^+\mu^-$ and $\tau^- \to e^- \mu^+\mu^-$ (shown in Fig.~\ref{Fig:tau3mu}). Concerning $\mu\to e\gamma$ we expect constraints which are at least comparable to the ones from $\mu^-\to e^-e^+e^-$ since $\mu\to e\gamma$ does not involve the small electron Yukawa coupling entering $\mu^-\to e^-e^+e^-$. In fact the constraints from $\mu\to e\gamma$ turn out to be stronger than the ones from $\mu^-\to e^-e^+e^-$ (see Fig.~\ref{Fig:muegamma}). Note that the constraints from $\mu^-\to e^-e^+e^-$ can be avoided if $v_u \epsilon^\ell_{11}\approx m_e$ while the leading contribution to $\mu\to e\gamma$ vanishes for $v_u \epsilon^\ell_{22}\approx m_\mu$. 
\medskip

\begin{figure}[t]
\centering
\includegraphics[width=0.44\textwidth]{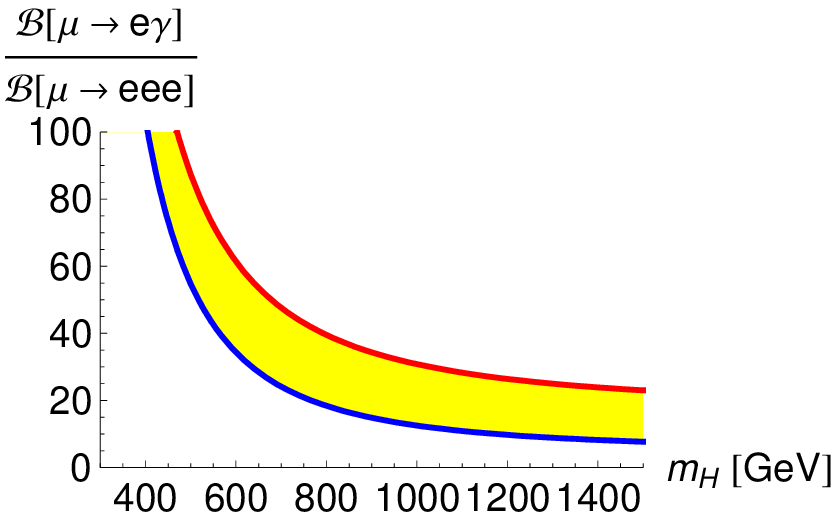}
\includegraphics[width=0.55\textwidth]{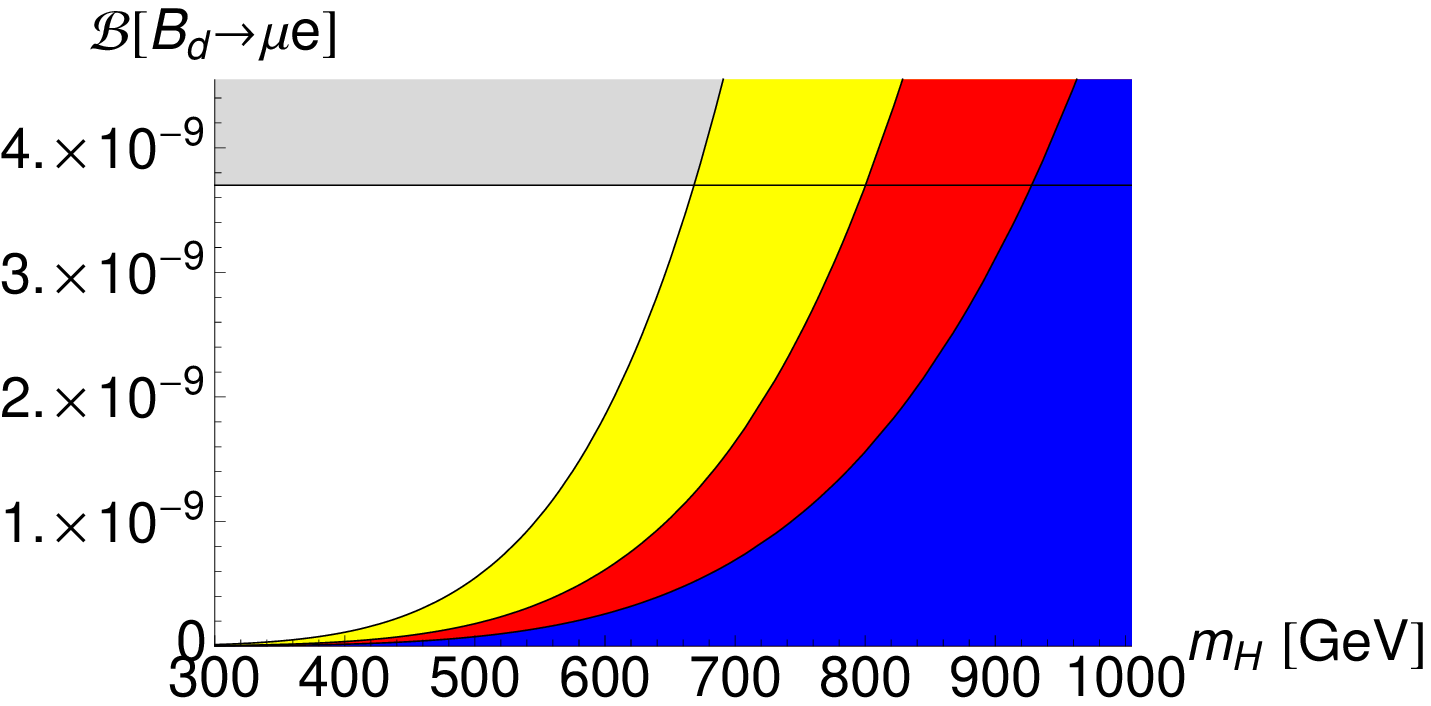}
\caption{Left: $\frac{{{\cal B} \left[ {{\mu} \to {e}\gamma } \right]}}{{{\cal B} \left[ {{\mu^{-}} \to {e^-}e^+ e^- } \right]}}$ as a function of $m_H$ assuming that only $\epsilon^\ell_{12}$ (red) or $\epsilon^\ell_{21}$ (blue) is different from zero for $\tan\beta=50$. For scenarios in which both $\epsilon^\ell_{12}$ and $\epsilon^\ell_{21}$ are different from zero the 2HDM of type III predicts the ratio $\left(\frac{{{\cal B} \left[ {{\mu} \to {e}\gamma } \right]}}{{{\cal B} \left[ {{\mu^{-}} \to {e^-}e^+ e^- } \right]}}\right)$ to be within the yellow region. These ratios are to a good approximation independent of $\tan\beta$ for $\tan\beta \gtrsim 20$. \newline
Right: Upper limits on the branching ratio of $B_d\to\mu e$ as a function of $m_H$ for $\tan\beta=30$ (blue), $\tan\beta=40$ (red) and $\tan\beta=50$ (yellow). The gray region is excluded by experiment.
\label{Fig:LFVmesondecays}}
\end{figure}

Finally, we can find correlations between lepton flavour-violating observables. Here we consider $\mu\to e$ transitions and ask the interested reader to inquire Ref.~\cite{Crivellin:2013wna} for $\tau\to \mu$ and $\tau\to e$ transitions. From the left plot of Fig.~\ref{Fig:LFVmesondecays} we see that there is a clear prediction for the ratio $\frac{{{\cal B} \left[ {{\mu} \to {e}\gamma } \right]}}{{{\cal B} \left[ {{\mu^{-}} \to {e^-}e^+ e^- } \right]}}$ in the 2HDM of type III. The right plot of Fig.~\ref{Fig:LFVmesondecays} shows the maximally allowed branching ratio for $B_d\to\mu e$ taking into account the constraints from $\mu\to e\gamma$ and $B_d\to\mu^+\mu^-$.


\begin{thebibliography}{99}

\bibitem{Crivellin:2013wna}
  A.~Crivellin, A.~Kokulu and C.~Greub,
  ``Flavor-phenomenology of two-Higgs-doublet models with generic Yukawa structure,''
  Phys.\ Rev.\ D {\bf 87} (2013) 094031
  [arXiv:1303.5877 [hep-ph]].

\bibitem{Crivellin:2012ye}
  A.~Crivellin, C.~Greub and A.~Kokulu,
  ``Explaining $B\to D\tau\nu$, $B\to D^*\tau\nu$ and $B\to \tau\nu$ in a 2HDM of type III,''  Phys.\ Rev.\ D {\bf 86} (2012) 054014  [arXiv:1206.2634 [hep-ph]].  

 
\bibitem{Lee:1973iz}
  T.~D.~Lee,
  ``A Theory of Spontaneous T Violation,''  Phys.\ Rev.\ D {\bf 8} (1973) 1226.  

\bibitem{Hamzaoui:1998nu}
  C.~Hamzaoui, M.~Pospelov and M.~Toharia,
  ``Higgs mediated FCNC in supersymmetric models with large $\tan \beta$,''  Phys.\ Rev.\ D {\bf 59} (1999) 095005  [hep-ph/9807350].  

\bibitem{Crivellin:2011jt}
  A.~Crivellin,
  ``Effective Higgs Vertices in the generic MSSM,''  Phys.\ Rev.\ D {\bf 83} (2011) 056001  [arXiv:1012.4840 [hep-ph]].  
  A.~Crivellin, L.~Hofer and J.~Rosiek,
  ``Complete resummation of chirally-enhanced loop-effects in the MSSM with non-minimal sources of flavor-violation,''  JHEP {\bf 1107} (2011) 017  [arXiv:1103.4272 [hep-ph]].  

\bibitem{Hermann:2012fc}
  T.~Hermann, M.~Misiak and M.~Steinhauser,
  ``$\bar{B}\to X_s \gamma$ in the Two Higgs Doublet Model up to Next-to-Next-to-Leading Order in QCD,''  JHEP {\bf 1211} (2012) 036  [arXiv:1208.2788 [hep-ph]].  

\bibitem{CMS}
  S.~Chatrchyan {\it et al.}  [CMS Collaboration],
  ``Search for supersymmetry in hadronic final states using MT2 in $pp$ collisions at $\sqrt{s} = 7$ TeV,''
  JHEP {\bf 1210} (2012) 018
  [arXiv:1207.1798 [hep-ex]].
	
\bibitem{Crivellin:2011ba}
  A.~Crivellin and L.~Mercolli,
  ``$B \to X_d \gamma$ and constraints on new physics,''
  Phys.\ Rev.\ D {\bf 84} (2011) 114005
  [arXiv:1106.5499 [hep-ph]].

\bibitem{BaBar:2012xj}
  J.~P.~Lees {\it et al.}  [BaBar Collaboration],
  ``Evidence for an excess of $\bar{B} \to D^{(*)} \tau^-\bar{\nu}_\tau$ decays,''  Phys.\ Rev.\ Lett.\  {\bf 109} (2012) 101802 [arXiv:1205.5442 [hep-ex]].  

\bibitem{Charles:2004jd}
  J.~Charles {\it et al.}  [CKMfitter Group Collaboration],
  ``CP violation and the CKM matrix: Assessing the impact of the asymmetric $B$ factories,''  Eur.\ Phys.\ J.\ C {\bf 41} (2005) 1  [hep-ph/0406184].  

\bibitem{Hou:1992sy}
  W.~-S.~Hou,
  ``Enhanced charged Higgs boson effects in $B \to \tau\nu$ and $B \to \tau\nu +X$,''  Phys.\ Rev.\ D {\bf 48} (1993) 2342.  
	  M.~Tanaka,
  ``Charged Higgs effects on exclusive semitauonic $B$ decays,''  Z.\ Phys.\ C {\bf 67} (1995) 321  [hep-ph/9411405].  

\bibitem{Tanaka:1994ay}
  M.~Tanaka,
  ``Charged Higgs effects on exclusive semitauonic $B$ decays,''
  Z.\ Phys.\ C {\bf 67} (1995) 321
  [hep-ph/9411405].

\bibitem{Crivellin:2009sd}
  A.~Crivellin,
  ``Effects of right-handed charged currents on the determinations of $|V_{ub}|$ and $|V_{cb}|$,''  Phys.\ Rev.\ D {\bf 81} (2010) 031301  [arXiv:0907.2461 [hep-ph]].  

 


\end{thebibliography}
\end{document}